\documentclass[preprint2,10.5pt]{aastex}
\begin{document}

\title{\large{\rm{WR~38/38a AND THE RATIO OF TOTAL-TO-SELECTIVE EXTINCTION IN CARINA}}}

\author{D.~G. Turner}
\affil{Saint Mary's University, Halifax, Nova Scotia, Canada}
\email{\rm{turner@ap.smu.ca}}

\begin{abstract}
A reanalysis of the (seemingly very distant) open cluster Shorlin~1, the group of stars associated with WR~38 and WR~38a, is made on the basis of existing {\it UBV} and {\it JHK}$_{\rm s}$ observations for cluster members. The 2MASS observations, in particular, imply a mean cluster reddening of $E_{B-V}=1.45\pm0.07$ and a distance of $2.94\pm0.12$ kpc. The reddening agrees with the {\it UBV} results provided that the local reddening slope is described by $E_{U-B}/E_{B-V}=0.64\pm0.01$, but the distance estimates in the 2MASS and {\it UBV} systems agree only if the ratio of total-to-selective extinction for the associated dust is $R=A_V/E_{B-V}=4.0\pm0.1$. Both results are similar to what has been obtained for adjacent clusters in the Eta Carinae region by similar analyses, which suggests that``anomalous'' dust extinction is widespread through the region, particularly for groups reddened by relatively nearby dust. Dust associated with the Eta Carinae complex itself appears to exhibit more ``normal'' qualities. The results have direct implications for the interpretation of distances to optical spiral arm indicators for the Galaxy at $\ell=287\degr-291\degr$, in particular the Carina arm here is probably little more than $\sim2$ kpc distant, rather than $2.5-3$ kpc distant as implied in previous studies.  Newly-derived intrinsic parameters for the two cluster Wolf-Rayet stars WR 38 (WC4) and WR 38a (WN5) are in good agreement with what is found for other WR stars in Galactic open clusters, which was not the case previously. 
\end{abstract}


\section{{\rm \footnotesize INTRODUCTION}}
\begin{figure}[t]
\begin{center}
\includegraphics[width=0.40\textwidth]{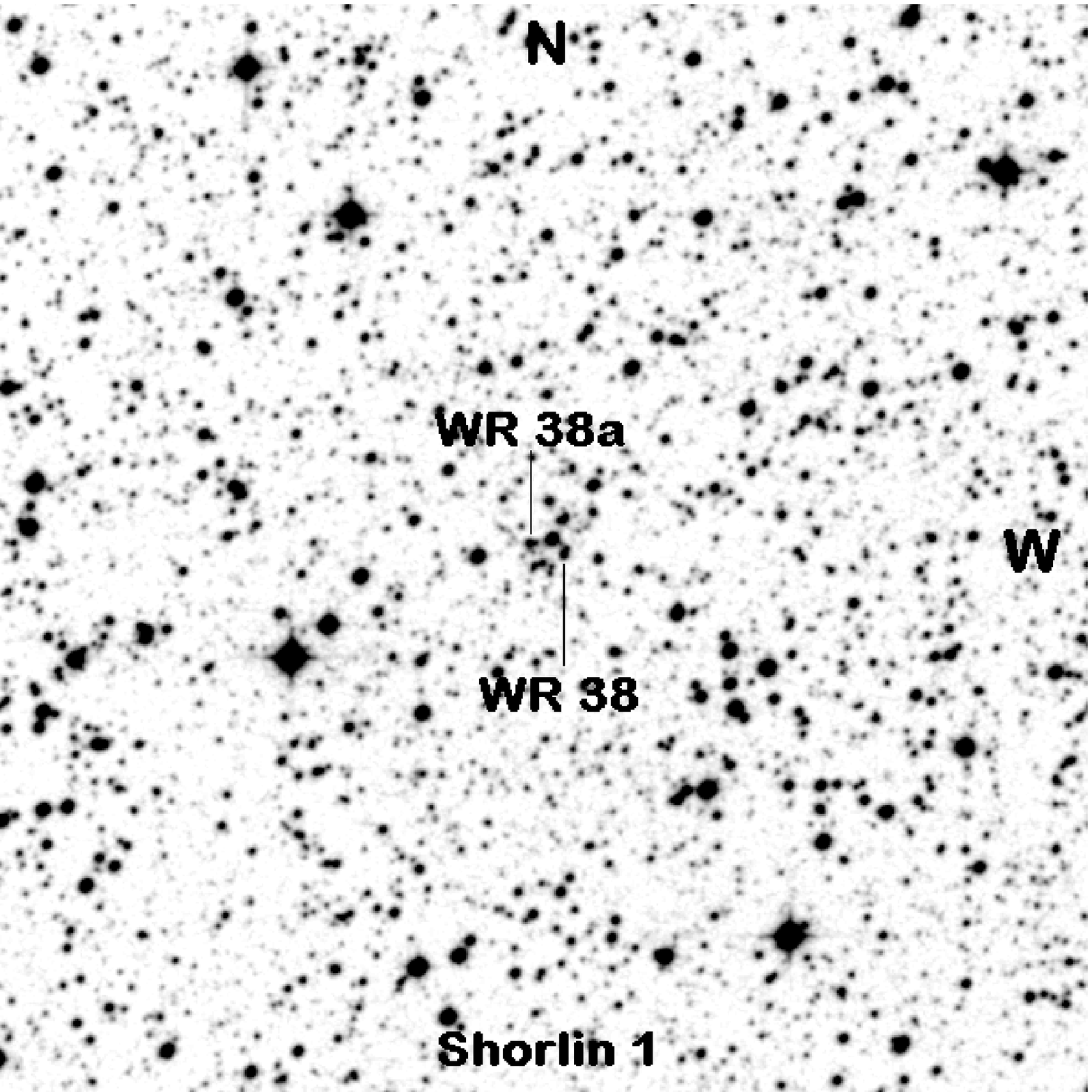}
\end{center}
\caption{\small{The $10\arcmin \times 10\arcmin$ field of Shorlin~1 and WR 38/38a from the ESO/SRC Sky Survey red image of the field. The image is centered on 2000 co-ordinates: 11:05:48, --61:13:14, which represents the core of the cluster Shorlin~1.}}
\label{fig1}
\end{figure}

The Galaxy's collection of open clusters is a varied assortment of stellar groups of diverse richness and age that serves a number of important functions for studies in Galactic astronomy, ranging from empirical investigations of the properties of interstellar extinction \citep{tu76b} and observational verification of computational models for stellar evolution \citep{me93} to the calibration of luminosities, interstellar reddenings, and ages for the sundry objects belonging to them \citep[e.g., Cepheids:][]{tu10}. Much of what we know about the intrinsic properties of Wolf-Rayet stars, for example, has been established with reference to the sample of such objects that are members of open clusters \citep[e.g.,][]{ls84,vh01}. The increased sensitivity of astronomical detectors in recent years has coincided with a growth in the number of Galactic calibrators found in open clusters, and has extended the range of such calibrators beyond the distance limits of previous years. One such case involves the pair of Wolf-Rayet stars WR 38 and WR 38a, which appear to be members of a distant open cluster \citep*{sh98,sh04,wa05}, recently designated as Shorlin~1 \citep[on-line update of][]{di02} and depicted in Fig.~\ref{fig1}.

A problem arises from the three photometric studies of Shorlin~1: a photoelectric/CCD {\it UBV} study by \citet{sh98} and \citet{sh04}, a {\it Hubble Space Telescope} (HST) {\it UBV} study by \citet{wa05} analyzed in conjunction with {\it JHK}$_{\rm s}$ observations \citep{cu03} from the {\it Two Micron All Sky Survey} \citep[2MASS,][]{sk06}, and a CCD {\it UBVI} study by \citet{cc09}. The study by \citet{sh04} implied a distance of $\sim 14.5$ kpc to Shorlin~1, the HST study by \citet{wa05} found a distance of $\sim 10$ kpc, and the study by \citet{cc09} a distance of $\sim12.7$ kpc. Observations from the 2MASS survey were included in the \citet{wa05} study, yet it is possible to use {\it JHK}$_{\rm s}$ observations independently to obtain estimates for the reddening and distance of the cluster \citep{tu11}. It was in an attempt to reconcile the various studies of Shorlin~1 by \citet{sh04}, \citet{wa05}, and \citet{cc09} by the techniques described by \citet{tu11} for analysis of 2MASS observations that a completely separate solution was obtained, one that has far-reaching implications for the delineation of Galactic spiral structure in the Carina region. That solution is described here.

\setcounter{table}{0}
\begin{table*}[t]
\caption{Estimates of {\it R} for the Carina Region.}
\label{tab1}
\centering
\small
\begin{tabular*}{0.98\textwidth}{@{\extracolsep{-0.1mm}}llll}
\hline \hline \noalign{\smallskip}
Field &$R\pm \Delta R$ &Method &Source \\
\noalign{\smallskip} \hline \noalign{\smallskip}
$\eta$~Carinae &$5.3\pm0.8$ &Comparison of radio/{\it H}$\alpha$ emission &\citet{rs67} \\
Tr~14/16 &$\sim 3$ &{\it UBV} VE study &\citet{fe69} \\
Car/Cen &$3.5\pm0.5$ &{\it UBV} VE study &\citet{lo72} \\
IC~2581 &$5.5\pm0.3$ &{\it UBV} VE study &\citet{tu73} \\
Tr~16 &$5.5\pm0.5$ &{\it UBV} VE study &\citet{tu74} \\
Car~R1 &$5.13\pm0.60$ &{\it UBV-}SpT VE study &\citet{he75} \\
HD~92964 &$3.2$ &Polarization $\lambda_{\rm max}$ &\citet{wh75} \\
IC~2581 &$5.3\pm0.2$ &{\it UBV} VE study &\citet{tu76a} \\
Car &$3.25\pm0.07$ &Mean polarization $\lambda_{\rm max}$ &\citet{wh77} \\
Tr~14/16/Cr~228 &$\sim 5$ &{\it UBVRI} VE study &\citet{he76} \\
IC~2581 &$3.11\pm0.18$ &{\it UBV-}SpT VE study &\citet{tu78} \\
Car~OB1 &$4.7\pm0.5$ &{\it BVRI} color excess ratios &\citet{fo78} \\
Carina,~$d<1$ kpc &$\sim 3.3$ &Mean polarization $\lambda_{\rm max}$ &\citet{wh79} \\
Tr~14/15/16/Cr~228/232 &$3.20\pm0.28$ &{\it UBV} VE study &\citet{tm80} \\
NGC~3293/3324 &$\sim 3.6$ &{\it UBVRI-}SpT VE study &\citet{te80} \\
Tr~15 &$\sim 3.2$ &{\it UBVRI} VE study &\citet*{fe80} \\
Tr~15/16/Cr~228 &$3.9\pm0.1$ &{\it UBVRIJHKLM} color differences &\citet*{th80} \\
Tr~16-149 &$4.4$ &{\it UBVRIJHKL}/Walraven colors &\citet{tg83} \\
Tr~14/16 &$4.7\pm0.1$ &{\it UBVRIJHK} color differences &\citet{sm87} \\
Tr~14/$\eta$ Car &$\sim 4$ &Nebula expansion/SpT distances &\citet{ah93,wa95} \\
Tr~14/16 &$\sim 3-4$ &SpT distances &\citet{wa95} \\
Tr~14 &$4.70\pm0.65$ &{\it UBVRI} VE study/color differences &\citet{va96} \\
Carina,~$d<1$ &$\sim 3.6$ &$E_{U-B}/E_{B-V}-R$ dependence &\citet{tu89,tu96} \\
Tr~14/15/16/Car~I &$>3.5$? &{\it UBVRIJHK} photometry &\citet{ta03} \\
NGC~3293 & $\sim 3.1$ &{\it UBVRI} color excess ratios &\citet{ba03} \\
Ru~91 &$3.82\pm0.13$ &{\it UBV} VE study &\citet{te05} \\
Cr~236 &$\sim3.82$ &{\it UBV-}SpT VE study &\citet{te09a} \\
Shorlin~1 &$4.0 \pm0.1$ &Comparison of {\it UBV}/{\it JHK}$_s$ distances &This paper \\
\noalign{\smallskip} \hline
\end{tabular*}
\end{table*}

The importance of the results for Shorlin~1 is tied to the ratio of total-to-selective extinction, $R=A_V/E_{B-V}$, relating the color excess $E_{B-V}$ to the total extinction in the Johnson visual {\it V} band, $A_V$. A knowledge of {\it R} is used to correct apparent distance moduli for reddened objects for the effects of interstellar extinction, and therefore plays an important role in establishing distances to such objects. The region surrounding the $\eta$ Carinae nebula has long been a controversial part of the Galactic plane in that respect, since a variety of studies over the years have generated a wide range of values for the value of {\it R} in Carina. Table~\ref{tab1} is an attempt to summarize the many and varied {\it R} analyses that have been carried out for stars and clusters in the Carina region, revealing that a range of solutions covering the interval from $R=3.1$ to $R=5.5$ has been generated, with no obvious consensus on the true value of {\it R} applying throughout the region. In fact, two studies \citep{wa95,ta03} argue that {\it R} has distinctly different values in the closely adjacent open clusters Tr~14 and Tr~16, while \citet{th80} have argued that it varies from one star to another in this direction. As pointed out here, both arguments are probably very close to the truth.

It should be noted in Table~\ref{tab1} that distinctly different methods have been employed to derive values of {\it R} describing the extinction in the designated fields of Carina. The abbreviation ``VE'' used in the table stands for the variable-extinction method, as described for example by \citet{tu76a,tu76b}. Each technique employed to derive {\it R} has its strengths and weaknesses, as discussed in \S3. The point is that applications of the same method often generate disparate values of {\it R}, and adoption of an overall mean for the region would only exacerbate the problem.

\section{{\rm \footnotesize THE OBSERVATIONAL DATA FOR SHORLIN~1}}

\begin{figure}[]
\begin{center}
\includegraphics[width=0.40\textwidth]{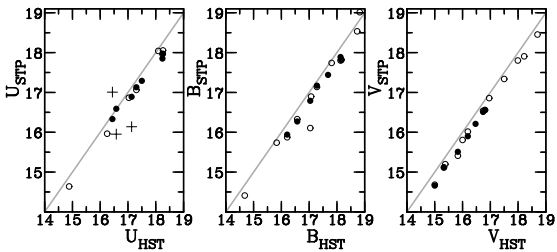}
\end{center}
\caption{\small{A comparison of the {\it UBV} magnitudes for stars in the studies of \citet{sh04} (STP, filled circles) and \citet{cc09} (open circles, plus signs) with those of \citet{wa05} (HST), with gray lines depicting the case for an exact match.}}
\label{fig2}
\end{figure}

Fig.~\ref{fig2} presents a comparison of the {\it UBV} observations of \citet{sh04} and \citet{cc09} with those of \citet{wa05} for stars in common. Since the \citet{cc09} observations were normalized to the \citet{sh04} data, they generally match the trends of the latter. Deviant points from the \citet{cc09} study (e.g., the plus signs in Fig.~\ref{fig2}) correspond to emission-line stars (WR 38 and WR 38a) and very blue stars, and were omitted from subsequent analysis. Clearly there are systematic offsets between data sets on the HST system and those normalized to the \citet{sh04} study, although the general trends are linear: 0.984 for {\it U}, 0.998 for {\it B}, and 1.016 for {\it V}. The ground-based CCD observations of \citet{sh04}, calibrated using relatively bright stars observed photoelectrically, might be the source of the discrepancy, particularly in {\it U} where a Balmer discontinuity correction was applied to the original observations with the CCD camera on the University of Toronto's Las Campanas 0.6m telescope. In contrast, the HST WFPC2 observations were tied to calibrations by \citet{ho95} and \citet{do00}, although there are indications that the sensitivity of the camera degrades temporally \citep{sa06}.

It is possible to adjust observations from one system to those of the other, so both systems were treated separately as the standard reference frame and the results examined individually. When the observations of \citet{sh04} and \citet{cc09} are standardized in such fashion to the HST {\it UBV} system of \citet{wa05}, the resulting data for likely cluster members lying within the cluster nucleus (see below) generate the color-color and color-magnitude diagrams of Fig.~\ref{fig3}. If the \citet{sh04} system is used as the standard reference frame, the {\it UBV} color-color and color-magnitude diagrams are as shown in Fig.~\ref{fig4}. The scatter in the observations is large in both cases, typically ranging between $\pm0^{\rm m}.10$ and $\pm0^{\rm m}.40$ in the colors. Although seemingly consistent with the cited uncertainties in the three studies, such large scatter may also be linked to differential reddening in the field, as suggested by the systematic tendency for faint stars to be bluer on average than brighter stars, the expected bias caused by interstellar extinction. It does limit what one can infer from the observations. The differences between Figs.~\ref{fig3}~and~\ref{fig4} are generally small, however, except for a magnitude offset, the \citet{sh04} system being brighter than that of \citet{wa05}.

\begin{figure}[t]
\begin{center}
\includegraphics[width=0.40\textwidth]{wrf3}
\end{center}
\caption{\small{{\it UBV} color-color (upper) and color-magnitude (lower) diagrams for Shorlin~1 from \citet{wa05} (filled circles) and transformed from \citet{sh04} (open circles) and \citet{cc09} (diamonds), with the data for WR 38 and WR 38a corrected for emission and shown by plus signs. Gray lines (upper) represent the intrinsic color-color relation for dwarfs, and the same relation reddened by $\pm0.06$ about a mean reddening of $E_{B-V}=1.46$ (black line). The solid black line and accompanying gray lines (lower) are ZAMS fits to the data for $E_{B-V}=1.46\pm0.06$ and {\it V--}$M_V=18.15\pm0.25$ (see text).}}
\label{fig3}
\end{figure}

\begin{figure}[t]
\begin{center}
\includegraphics[width=0.40\textwidth]{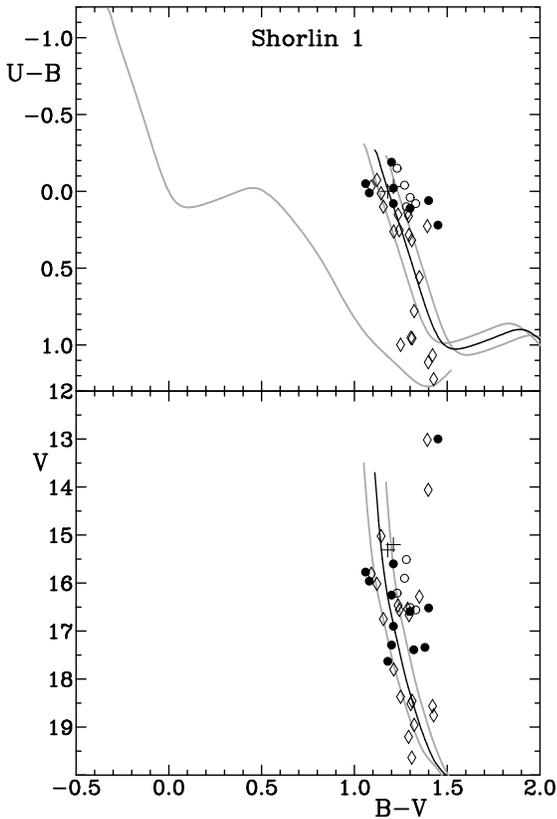}
\end{center}
\caption{\small{The same as Fig.~\ref{fig3} with the data transformed to the \citet{sh04} system and for $E_{B-V}=1.44\pm0.06$ and {\it V--}$M_V=18.20\pm0.20$.}}
\label{fig4}
\end{figure}

The {\it UBV} observations of Figs.~\ref{fig3}~and~\ref{fig4} can be analyzed by standard techniques \citep[e.g.,][]{tu76a,te80,te05}, but an essential first step is to confirm the reddening relation for the field \citep{tu89}. If one makes the unjustified assumption that the reddening in Shorlin~1 is typical of that in other regions of the Galaxy, then one finds that adoption of a mean relation such as $E_{U-B}=0.72E_{B-V}+0.02E_{B-V}^2$ \citep{tu89} results in 5 stars bluer than the intrinsic colors for hot O7 stars. Adoption of a mean relation such as $E_{U-B}=0.72E_{B-V}+0.05E_{B-V}^2$ \citep{fi70} is worse, since it results in 11 stars bluer than the intrinsic colors of O7 stars. Most clusters containing early-type Wolf-Rayet stars as members also contain evolved O-type stars and B supergiants \citep[see][]{ls84}, so it seems unlikely that Shorlin~1 could be so rich in early O-type stars.

Alternatively, one can establish the appropriate reddening slope for the region of Shorlin~1 from the colors of the bluest member stars. The sample of blue stars is small, but there are three objects that have colors possibly matching those of a reddened O7 star (2 objects) or B1 star (1 object). The implied reddening slope for the region is then $E_{U-B}/E_{B-V}=0.64(\pm0.01) +0.02E_{B-V}^2$, which was adopted here. There are no spectroscopic observations available to test such an assumption, but one is free to adopt later spectral types for the stars, in which case the implied reddening slope becomes even smaller (also a possibility).

The analysis of Shorlin~1 stars proceeded as follows. The intrinsic color-color relation for dwarfs was reddened by various amounts to match the observed colors of cluster stars, and the intrinsic color-absolute magnitude relation for zero-age main-sequence stars \citep[ZAMS,][]{tu76a,tu79} was adjusted simultaneously in magnitude to examine the fit for obvious main-sequence stars. A minimum reddening was found by matching the color-color diagram and color-magnitude diagram to the bluest cluster stars, and a maximum reddening was found in similar fashion using the reddest cluster stars. Several stars lay beyond such limits, but in such small numbers that a simultaneous fit was impossible. That may be yet further evidence for the possible existence of differential reddening in the field. The results were similar for both Fig.~\ref{fig3} and Fig.~\ref{fig4}, yielding values of $E_{B-V}=1.46\pm0.06$ and {\it V--}$M_V=18.15\pm0.25$ for data on the HST system, and very similar values of $E_{B-V}=1.44\pm0.06$ and {\it V--}$M_V=18.20\pm0.20$ for data on the system of \citet{sh04}. The cluster color-magnitude diagrams of Figs.~\ref{fig3}~and~\ref{fig4} also appear to contain a few evolved blue supergiants, typical of groups containing Wolf-Rayet stars \citep{ls84}. Such stars were, in fact, suspected in the original study by \citet{sh04}.

The standard reddening relation for nearby dust clouds in the Galactic plane has a slope that averages 0.72, but does range from values as small as 0.54--0.62 to values of 0.80 or more \citep{jm55,tu89}. There are also indications in nearby regions of the Galaxy for a correlation between {\it UBV} reddening slope and the ratio of total-to-selective extinction, $R=A_V/E_{B-V}$, describing the extinction properties of the dust \citep{tu94,tu96}. The average size of dust grains in the Galactic plane affects the optical reddening slope $E_{U-B}/E_{B-V}$ in consistent fashion with how the ratio of total-to-selective extinction {\it R} is affected, namely small reddening slopes are linked to larger-than-average values of {\it R} and large reddening slopes are linked to smaller-than-average values of {\it R}. That is expected if the grain size distribution for the dust particles is skewed to larger-than-average dimensions in the former situation, and to smaller-than-average dimensions in the latter situation. Nor are the peculiarities of dust extinction in the direction of Shorlin~1 ($\ell=290\degr.63$) unique, namely a smaller-than-average reddening slope. The extinction associated with the region of Ruprecht 91 at $\ell=286\degr.87$ is described by a reddening slope of $E_{U-B}/E_{B-V}=0.65\pm0.02$ and $R=3.82\pm0.13$ \citep{te05}, and similar results with $E_{U-B}/E_{B-V}=0.64\pm0.13$ apply to the dust extinction in Collinder 236 at $\ell=289\degr.58$ \citep{te09a}. The results for Shorlin~1 are therefore consistent with previous results for the extinction properties of dust affecting clusters in the Carina region.

The value of the reddening slope is important for previous analyses of the cluster, since the study by \citet{sh04} used previous results for the reddening slope and value of {\it R} in Carina \citep{tu76b,tu77,tu78,tm80,te80} to infer the reddening and distance of Shorlin~1. A reddening slope of $E_{U-B}/E_{B-V}=0.64$ for Shorlin~1 results in a significantly smaller mean cluster reddening ($E_{B-V}=1.45\pm0.07$), as noted above, than the value of $E_{B-V}=1.60$ obtained by \citet{sh04}. And a correspondingly different (presumably larger) value for {\it R} results in a much smaller derived distance to the cluster than the published values ranging from $10-14.5$ kpc. The exact amount of change depends upon the value of {\it R} for Shorlin~1.

\begin{figure}[t]
\begin{center}
\includegraphics[width=0.40\textwidth]{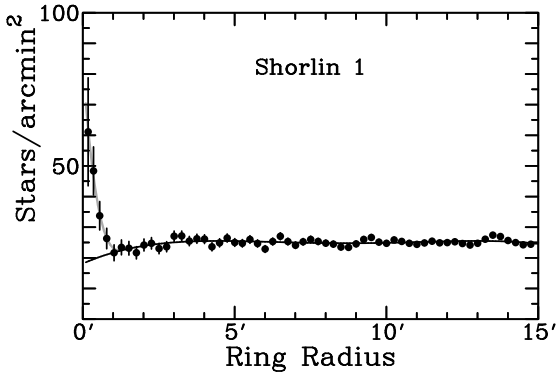}
\end{center}
\caption{\small{Star counts in the field of Shorlin~1 about the adopted cluster center in Fig.~\ref{fig1}, with Poisson uncertainties included with each point. The solid black line depicts the variation in background star densities across the field, and a gray line shows a best fit to cluster core region stars.}}
\label{fig5}
\end{figure}

An alternate approach to the problem can be made using 2MASS observations of cluster stars \citep{cu03}. Intrinsic relations for {\it JHK}$_{\rm s}$ photometry are available from \citet{tu11}, and are linked to the same intrinsic {\it UBV} colors for main sequence stars and the same zero-age main-sequence relation \citep{tu76a,tu79}. A first step is to restrict the analysis to likely cluster members. For that a cluster center of symmetry at 2000 co-ordinates 11:05:48, --61:13:44, was identified by eye from Fig.~\ref{fig1}, and star counts were made to the limit of the 2MASS survey in {\it J} in $0\arcmin.25$ rings about that point. The results are shown in Fig.~\ref{fig5} along with Poisson uncertainties for each datum.

\begin{figure}[t]
\begin{center}
\includegraphics[width=0.40\textwidth]{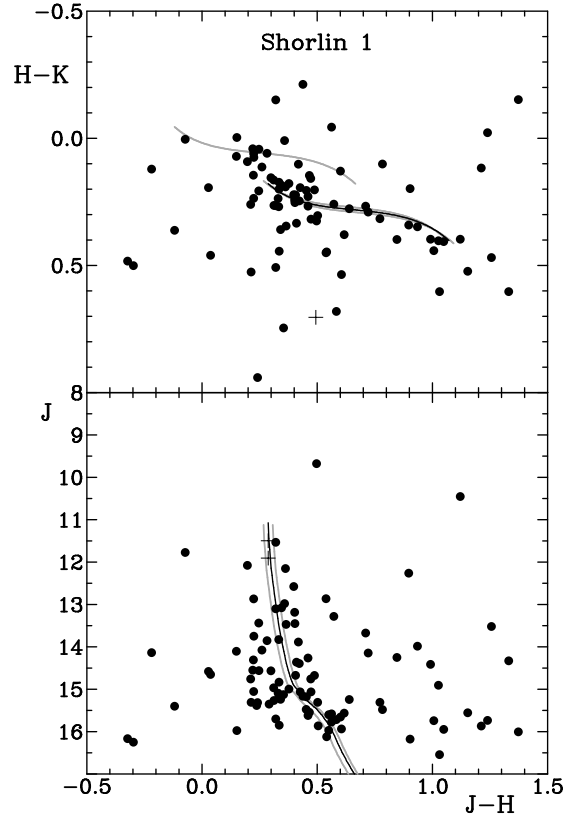}
\end{center}
\caption{\small{{\it JHK}$_{\rm s}$ color-color (upper) and color-magnitude (lower) diagrams for Shorlin~1 from the data of \citet{cu03}. Gray lines correspond to the intrinsic color-color relation for dwarfs (upper) and the extrema of reddening from Figs.~\ref{fig3}~and~\ref{fig4} (upper and lower) converted to the 2MASS system; black lines correspond to $E_{B-V}=1.45\pm0.07$ and {\it V--M}$_V=12.31$, {\it E(J--H)} $=0.406\pm0.020$ and {\it J--M}$_J=13.40$. Crosses represent data for the two Wolf-Rayet stars WR 38 and WR 38a, where, because of anomalies in the {\it JHK}$_{\rm s}$ colors, the {\it J--H} colors for both stars in the lower diagram are taken to be identical to reddened O-type cluster members.}}
\label{fig6}
\end{figure}

Star counts for Shorlin~1 appear to differ in one important property relative to those for most other open clusters: they do not exhibit the core/halo (nucleus/corona) structure typical of other clusters \citep{kh69,te05,te09a}. The implication is that Shorlin~1 does not exhibit the characteristics of a true star cluster, but more closely resembles a stellar asterism, perhaps the compact trapezium-like core of a now-dissolved young cluster. The star densities do exhibit a peak at the adopted cluster center, but immediately outside the (assumed) cluster nucleus decrease to values that are smaller than in more distant regions, as if the total extinction along the line of sight increased slightly near the center of the field. Otherwise the field star density in the region is reasonably constant. If the radial trend follows the relation shown in Fig.~\ref{fig5}, then the core of the dissolved cluster has a radius of $\sim1\arcmin.25$, within which there are $\sim 32$ Shorlin~1 members brighter than the 2MASS survey limit of $J\simeq 17$ (an uncertainty cannot be established because of the {\it ad hoc} assumption about background star densities). Since likely Shorlin~1 members lie within $1\arcmin$ of the adopted group center, {\it JHK}$_{\rm s}$ color-color and color-magnitude diagrams for such stars were examined in conjunction with the {\it UBV} color-color and color-magnitude diagrams, as shown in Fig.~\ref{fig6}.

\setcounter{table}{1}
\begin{table*}[!ht]
\caption{New Parameters for Carina Clusters.}
\label{tab2}
\centering
\normalsize
\begin{tabular*}{0.840\textwidth}{@{\extracolsep{+0.5mm}}lcccl}
\hline \hline \noalign{\smallskip}
Cluster &$E_{B-V}$ &$V_0-M_V$ &{\it d} (kpc) &Source \\
\noalign{\smallskip} \hline \noalign{\smallskip}
Tr~14 &$0.71 \pm0.11$ &$11.69 \pm0.70$ &$2.17 \pm0.63$ &This paper \\
$\cdots$ &$0.33-0.74$ &$12.50 \pm0.20$ &$3.10 \pm0.30$ &\citet{va96} \\
Tr~15 &$0.54 \pm0.11$ &$11.56 \pm0.55$ &$2.06 \pm0.45$ &This paper \\
$\cdots$ &$0.48 \pm0.07$ &$12.10 \pm0.15$ &$2.60 \pm0.20$ &\citet{fe80} \\
Tr~16 &$0.73 \pm0.16$ &$11.57 \pm0.80$ &$2.06 \pm0.66$ &This paper \\
$\cdots$ &$0.38-0.80$ &$12.65 \pm0.20$ &$3.39 \pm0.30$ &\citet*{fe73} \\
NGC~3293 &$0.20 \pm0.05$ &$11.76 \pm0.20$ &$2.25 \pm0.17$ &This paper \\
$\cdots$ &$0.20-0.62$ &$11.99 \pm0.13$ &$2.50 \pm0.15$ &\citet{te80} \\
NGC~3324 &$0.43 \pm0.11$ &$11.59 \pm0.30$ &$2.08 \pm0.21$ &This paper \\
$\cdots$ &$0.47 \pm0.08$ &$12.47 \pm0.22$ &$3.12 \pm0.22$ &\citet{cl77} \\
IC~2581 &$0.43 \pm0.07$ &$11.59 \pm0.30$ &$2.08 \pm0.24$ &This paper \\
$\cdots$ &$0.457 \pm0.08$ &$12.29 \pm0.15$ &$2.87 \pm0.20$ &\citet{tu78} \\
Cr~228 &$0.21 \pm0.11$ &$11.45 \pm0.30$ &$1.95 \pm0.20$ &This paper \\
$\cdots$ &$0.18-0.49$ &$12.00 \pm0.20$ &$2.51 \pm0.23$ &\citet{fe76} \\
Cr~232 &$0.23 \pm0.09$ &$9.98 \pm0.30$ &$0.99 \pm0.08$ &This paper \\
Average & & &$2.09 \pm0.09$ &This paper \\
\noalign{\smallskip} \hline
\end{tabular*}
\end{table*}

Variations in optical extinction slope $E_{U-B}/E_{B-V}$ and $R=A_V/E_{B-V}$ affect reddening slope and total extinction corrections in the infrared at wavelengths of 1.235 ({\it J}), 1.662 ({\it H}), and 2.159 $\mu$m ({\it K}$_{\rm s}$) to only a minor extent. It is therefore appropriate to use standard techniques to analyze the observations of Fig.~\ref{fig6}. The {\it JHK}$_s$ observations were used in conjunction with the reddenings obtained for group stars in Figs.~\ref{fig3}~and~\ref{fig4}, with the results shown in Fig.~\ref{fig6}. The average reddening for Shorlin~1 stars obtained with the HST \citep{wa05} and \citet{sh04} scales corresponds on the 2MASS system to {\it E(J--H) =}$0.406\pm0.020$, which provides a reasonable fit to the observed {\it JHK}$_s$ data. The corresponding fit in distance modulus is {\it J--M}$_J=13.40$, with very little scatter. Most of the uncertainty originates in the implied reddening. The implied intrinsic distance modulus for Shorlin~1 is $J_0${\it --M}$_J=12.34\pm0.09$, corresponding to a distance of $2.94\pm0.12$ kpc, a distance considerably smaller than the values obtained by \citet{sh04}, \citet{wa05}, and \citet{cc09}.

The {\it UBV} data are consistent with the {\it JHK}$_{\rm s}$ results only if {\it R} is larger than traditional values near 3.1 \citep{tu76b}. The apparent distance moduli from ZAMS fitting in {\it UBV} and {\it JHK}$_s$ coincide only if $R=4.0\pm0.1$. The good agreement of that result with the value of $R=3.82\pm0.13$ obtained independently from a variable-extinction analysis by \citet{te05} for the extinction in nearby Ruprecht 91 suggests consistent (and ``anomalous'') properties for the dust extinction across much of the Carina region, as also argued in many of the studies summarized in Table~\ref{tab1}.

\section{{\rm \footnotesize THE EXTINCTION LAW IN CARINA}}
Some caution is required when interpreting the results of Table~\ref{tab1}, since each method used to derive {\it R} is subject to different sources of bias. The color difference method, for example, is susceptible to possible circumstellar emission, typical of luminous stars, which can produce excess emission in the far infrared, as well as to possible systematic effects arising from comparison of observed infrared colors with adopted intrinsic colors for the stars \citep[e.g.,][]{tu94}. The variable-extinction (VE) method, which by contrast is tied to ZAMS fitting for open clusters, is unaffected by circumstellar effects in stars. However, it can be biased by systematic and random errors in the photometry \citep{tu76a} and by limited ranges of color excess $E_{B-V}$ for clusters studied, which unfortunately is the case for most clusters in the Carina region. {\it UBV} photometry also suffers from possible systematic offsets in {\it U--B} colors arising from the use of ``non-standard'' photometers and filter mismatches relative to the Johnson system \citep{mv77} and from the observational treatment of atmospheric extinction \citep{cc01}.

The wavelength of maximum interstellar polarization is related directly to mean particle size along the line of sight, and hence {\it R-}value, but can be biased by relatively nearby dust and what is known about its extinction properties \citep{wh79}. The same problem arises with the use of color excess ratios. Do specific results apply to the dust producing variable extinction in the field, or to relatively nearby dust? The comparison of radio emission from the Carina nebula with Balmer line emission is reasonably straighforward, although concerns arise regarding whether or not the emission in the two wavelength regions can be treated in simple fashion, without regard to the patchy nature of the emission at visible wavelengths. In summary, many of the results presented in Table~\ref{tab1} may be quite reliable; others may not. A reasonable question to ask is: what circumstances would permit most of the results to be reliable?

An important consequence of the present results for Shorlin~1 is that they lead to specific predictions regarding the properties of dust extinction towards Carina. First, the possibility that $R>3$ throughout Carina implies that distances to clusters in the Eta Carinae region may all be overestimated, given that most studies have adopted a value of $R\simeq3$ to correct for the effects of extinction on apparent distance moduli of the clusters. Second, since variations in {\it R} are invariably tied to color excess ratios describing the dust extinction \citep{tu94,tu96}, it is reasonable to expect unusual results for the reddening slope towards stars in Carina clusters. The latter complication is one reason why a previous study of {\it R-}values for open clusters in the Galaxy neglected most open clusters in the Carina region \citep{tu76b}.

\begin{figure}[!t]
\begin{center}
\includegraphics[width=0.40\textwidth]{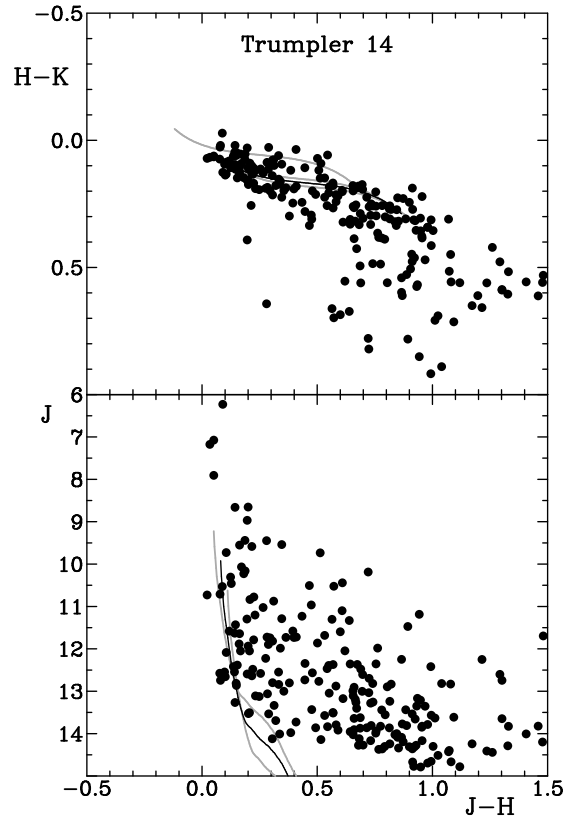}
\end{center}
\caption{\small{{\it JHK}$_{\rm s}$ color-color (upper) and color-magnitude (lower) diagrams for Trumpler~14 from the data of \citet{cu03}, only for stars with magnitude uncertainties smaller than $\pm0.05$. Features are identical to those of Fig.~\ref{fig6}, but for {\it E(J--H)} $=0.20\pm0.03$ and {\it J--M}$_J=12.20\pm0.70$, corresponding to $E_{B-V}=0.71\pm0.11$ and {\it V$_0$--M}$_V=11.69\pm0.70$.}}
\label{fig7}
\end{figure}

In order to test for the former effect, new estimates of distance and reddening were derived for open clusters in the Carina region using 2MASS photometry for the analysis, as done previously for a variety of open clusters \citep{te09a,te09b,tu11} and in this paper for Shorlin~1. Results of such an analysis are presented in Table~\ref{tab2}, and illustrated in Fig.~\ref{fig7} for the example of Trumpler 14 (one of the eight clusters studied). In most cases the analysis was restricted to stars in the innermost $2\arcmin-3\arcmin$ of the cluster center, usually available from the WEBDA on-line open cluster data base. The coordinates for Cr~228, however, were those of \citet{tm80}, i.e. 10:44:00, --60:05:00 (J2000), rather than what is cited in WEBDA. Also, Cr~232 is often considered to be linked with Tr~14 and Tr~16 in the literature, but in this study the cluster was taken to be the group of bright stars delineated by \citet{tm80}. Fainter stars in the cluster region may be mainly coronal members of Tr~14 and Tr~16, as also indicated by the proper motion survey of \citet*{cu93}, which implies that bright members of Cr~232 may be only $\sim 1$ kpc distant, as found here.

The newly-derived distances to open clusters in the Eta Carinae field, with the exception of Cr~232, all cluster around 2.1 kpc, the formal mean being $2.09\pm0.09$ kpc, as noted in Table~\ref{tab2}. The literature estimates of distance to the same clusters, selected references for which are listed in Table~\ref{tab2}, are consistently larger, as expected for studies in which $R\simeq3$ was used to correct the apparent distance moduli for the effects of extinction. A more suitable comparison can be made with the estimated distance to Eta Carinae itself, for which nebular expansion parallaxes have been generated in recent years. Studies by \citet{ah93} and \citet{sm06} led to estimated distances of $2.2\pm0.2$ kpc and $2.35\pm0.05$ kpc, respectively, for the distance to Eta Carinae from the geometry of the bipolar expansion of the Homunculus nebula around the star. \citet{sm08} summarize that as a distance of 2.3 kpc, accurate to $\pm2\%$. The good agreement with the {\it JHK}$_s$ distances to the star clusters in the Eta Carinae complex, particularly for the adjacent clusters Tr~14 and Tr~16, is what one expects if {\it R} is indeed anomalous in the region.

\begin{figure}[t]
\begin{center}
\includegraphics[width=0.40\textwidth]{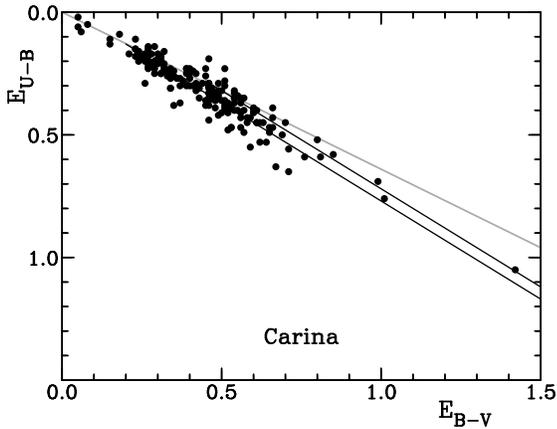}
\end{center}
\caption{\small{Derived color excesses for spectroscopically-observed stars in NGC~3324, NGC~3293, Tr~14, Tr~15, Tr~16, IC~2581, and Cr~228. The gray line represents a reddening relation of slope $E_{U-B}/E_{B-V}=0.64$, while black lines represent reddening relations of slope $E_{U-B}/E_{B-V}=0.80$ applying to the former after zero-point offsets of $E_{B-V}=0.20$ and 0.45.}}
\label{fig8}
\end{figure}

A second effect should be seen in the color excesses for stars populating the region. The methodology of \citet{tu89} was applied here to stars belonging to the Carina clusters NGC~3324, NGC~3293, Tr~14, Tr~15, Tr~16, IC~2581, and Cr~228, where the spectral types and photoelectric {\it UBV} observations used in the analysis were taken from the WEBDA on-line data base. The results are displayed in Fig.~\ref{fig8}. The scatter here is somewhat larger than that exhibited by the smaller starfields analyzed for the 1989 study, possibly because of errors in the assigned spectral types of some stars, but does lead to a reasonable interpretation. For stars of small color excess, and even for some of larger reddening, there is a reasonably good fit to a reddening slope of $E_{U-B}/E_{B-V}=0.64$, whereas stars of large reddening are more reasonably matched to a reddening slope of $E_{U-B}/E_{B-V}\simeq0.80$, where the changeover from the former to the latter has been arbitrarily set at $E_{B-V}=0.20$ and 0.45 to provide reasonable eye fits to the data.

The nature of the trends in the data of Fig.~\ref{fig8} is that the foreground extinction is anomalous, and is described by $E_{U-B}/E_{B-V}=0.64$ and $R=A_V/E_{B-V}=4.0$, whereas extinction arising in dust closer to the Eta Carinae complex itself is described by more ``normal'' extinction, i.e. $E_{U-B}/E_{B-V}\simeq0.80$ and $R=A_V/E_{B-V}\simeq3.0$. The changeover occurs at different reddenings along individual lines of sight, presumably as a result of variations in optical depth for the anomalous foreground dust extinction, and may account for the individualistic estimates of {\it R} for separate cluster stars in some studies \citep{th80} as well as for the unusual scatter of the data points in Fig.~\ref{fig8}. Apparently the ``anomalous'' dust in Carina is located foreground to the Carina arm, which must now be considered to be only $\sim2$ kpc distant in this direction, rather than $\sim2.5-3$ kpc distant as established previously.

In similar fashion, the group of stars associated with WR~38 and WR~38a, and referred to as Shorlin~1, must lie near the inside edge of the Carina-Sagittarius spiral arm, not well beyond it, given that the present study places it at a reduced distance of only $\sim2.9$ kpc. An independent test of the spectroscopic data was made using the CCD {\it UBV} photometry of \citet{ma93} for a few of the clusters. The resulting scatter was only marginally greater than that in Fig.~\ref{fig8}, and confirmed the results of the latter, but the sample size was smaller and therefore less useful to the analysis.

\section{{\rm \footnotesize DISCUSSION}}
The newly-derived distance of $2.94\pm0.12$ kpc to Shorlin~1, which makes use of 2MASS {\it JHK}$_s$ observations of cluster stars, is significantly smaller than previous estimates that placed it $10-14.5$ kpc from the Sun \citep{sh98,sh04,wa05,cc09}. That is a consequence of ``anomalous'' extinction by local dust clouds, mainly within a kiloparsec, characterized by a ratio of total-to-selective extinction of $R=4.0\pm0.1$. The larger distances estimates in previous studies are a systematic effect arising from analyzing the optical photometry for this heavily-reddened group using standard reddening and extinction relations. A small contradiction arises, given that the small dimensions of the cluster were used by \citet{sh04} as evidence for extreme distance. The implied cluster radius of $1\arcmin.25$ from star counts (Fig.~\ref{fig4}) corresponds to 1.1 pc at a distance of 2.94 kpc, less than half as large as the values that are typical of older, dynamically-evolved, open clusters \citep[e.g.,][]{wi71}. But the star counts also imply that Shorlin~1 is unlikely to represent a true cluster, but is instead merely the trapezium-like remains of a previously-bound cluster with an implied age of only a few Myr, according to the likely presence of O-type stars within its boundaries.

The reddening implied for cluster members is reduced from $E_{B-V}= 1.60$ to $E_{B-V}= 1.45\pm0.07$ with a shallow reddening slope of $E_{U-B}/E_{B-V}=0.64\pm0.01$, a value consistent with the larger-than-average value of {\it R} applying to foreground dust throughout the Carina field. A value of $R\simeq4$ for foreground dust of varying optical depths along most sight lines in the Carina region, in combination with dust extinction with more ``normal'' parameters within the Carina arm itself, can account for the unusual distribution of $E_{U-B}$ and $E_{B-V}$ color excesses for early-type stars in Carina clusters (Fig.~\ref{fig8}), as well as the unusual spread of {\it R}-estimates for Carina found in previous studies (e.g., Table~\ref{tab1}). A consequence of the anomalous foreground extinction is that the distances to other star clusters in the Eta Carinae complex are reduced from $\sim2.5-3$ kpc to values ranging from $\sim 2$ kpc to $\sim 2.3$ kpc, the distance found for Eta Carinae itself from its nebular expansion parallax \citep{ah93,sm06,sm08}. That clearly affects the interpretation of distances to other optical spiral arm features in the region. A case in point is the presently-inferred location of Shorlin~1 on the inside edge of the Sagittarius-Carina spiral arm according to its new distance, relative to a much more distant inferred location in an extension of the local or Perseus arms resulting from the studies by \citet{sh04} and \citet{cc09}. Similar changes affect conclusions by \citet{sh91} about the location of Wolf-Rayet stars in Carina and by \citet{mo02} and \citet{wa05} regarding the distance to the starburst cluster NGC 3603, which lies closely adjacent to the field of Shorlin~1.

\setcounter{table}{2}
\begin{table}[t]
\caption{Inferred Parameters for WR 38 and WR 38a.}
\label{tab3}
\centering
\begin{tabular}{{lcc}}
\hline \hline \noalign{\smallskip}
Parameter &WR 38 &WR 38a \\
\noalign{\smallskip} \hline \noalign{\smallskip}
Spectral Type &WC4 &WN5 \\
{\it v} &+15.60 &+16.21 \\
{\it b--v} &+0.88 &+0.83 \\
{\it u--b} &+0.60 &$\cdots$ \\
({\it b--v})$_0$ &--0.32 &--0.37  \\
\vspace{0.05in}
{\it M}$_v$ &--3.14 &--2.53  \\
{\it V}(HST)$_{\rm adj}$ &+15.45 &+15.55 \\
({\it B--V})(HST)$_{\rm adj}$ &+1.20 &+1.18 \\
({\it U--B})(HST)$_{\rm adj}$ &--0.04 &+0.01 \\
({\it B--V})(HST)$_0$ &--0.26 &--0.24  \\
\vspace{0.05in}
{\it M}(HST)$_V$ &--2.73 &--2.63  \\
{\it V}(STP)$_{\rm adj}$ &+15.20 &+15.31 \\
({\it B--V})(STP)$_{\rm adj}$ &+1.21 &+1.18 \\
({\it U--B})(STP)$_{\rm adj}$ &--0.03 &+0.00 \\
({\it B--V})(STP)$_0$ &--0.23 &--0.27  \\
{\it M}(STP)$_V$ &--2.90 &--2.79  \\
\noalign{\smallskip} \hline
\end{tabular}
\end{table}

The implied properties of the two Wolf-Rayet stars are also affected by the changes, primarily by the new reddening since {\it R}-values affect derived distances directly but have only a minor influence on luminosities derived from ZAMS fitting. If the same type of analysis is followed as in the study of \citet{sh04}, then the narrow band {\it ubv} photometry of the two stars by \citet{tm88}, \citet*{se90}, and \citet{sh91}, and the combined broad band {\it UBV} photometry of \citet{sh04}, \citet{wa05}, and \citet{cc09}, generate the parameters given in Table~\ref{tab3}, where the {\it UBV} values have been modified in approximate fashion to account for emission-line contamination. That was done by examining the narrow-band {\it ubv} observations of the Wolf-Rayet stars corrected for emission-line contamination and then adjusted to equivalent broad band {\it UBV} values with the relationships of \citet{tu82}, and comparing the resulting corrections $\Delta V$, $\Delta(B-V)$, and $\Delta(U-B)$ with those derived by \citet{py66} for WC4 and WN5 stars. Small differences were found relative to the \citet{py66} corrections, the best values for the corrections used here being +0.50, --0.02, and --0.40, respectively, for WC4, and +0.20, --0.02, and --0.20, respectively, for WN5. The adjusted values are cited separately for the \citet{sh04} scale (STP) and the \citet{wa05} scale (HST). It is stressed that the values in Table~\ref{tab3} are of rather low precision, and a thorough, deep, photometric survey of Shorlin~1 and the two Wolf-Rayet stars should help to improve both the precision and accuracy of the cited values.

The equivalent reddening for $E_{B-V}=1.45\pm0.07$ in the narrow band system is $E_{b-v}=1.20\pm0.06$, and generates the values given in the top portion of Table~\ref{tab3}. The broad band magnitudes and colors in the lower portion were corrected according to the results derived previously in Figs.~\ref{fig3}~and~\ref{fig4}. The new values for the intrinsic parameters of the two Wolf-Rayet stars, in particular the implied intrinsic colors and visual luminosities near $M_V\simeq M_v\simeq-3$, are now in reasonably good agreement with parameters derived for Wolf-Rayet stars of similar ionization class in open clusters \citep[see][]{vh01,tu82,ls84}, rather than being too blue and too luminous, as was previously the case when the stars were analyzed using extinction relations more typical of the Galactic mean \citep{sh04}. That would seem to be further support for the revised analysis of Shorlin~1 presented here.

\subsection*{{\rm \scriptsize ACKNOWLEDGEMENTS}}
\small{This publication makes use of data products from the Two Micron All Sky Survey, which is a joint project of the University of Massachusetts and the Infrared Processing and Analysis Center/California Institute of Technology, funded by the National Aeronautics and Space Administration and the National Science Foundation.}

\end{document}